\documentclass[conference]{IEEEtran}
\IEEEoverridecommandlockouts
\usepackage{cite}
\usepackage{amsmath,amssymb,amsfonts}
\usepackage{algorithm}
\usepackage{algpseudocode}
\usepackage{geometry}
\usepackage{graphicx}
\usepackage[footnotesize]{caption}
\usepackage{subcaption}  
\usepackage{textcomp}
\usepackage{tikz}
\usepackage{multirow}
\usepackage{hyperref}
\usetikzlibrary{positioning, shapes, arrows.meta}
\usepackage{standalone}
\usepackage{comment}
\usepackage{xcolor}
\usepackage{svg}

\makeatletter
\renewcommand{\section}{\@startsection{section}{1}{\z@}%
  {1.5ex plus 1.5ex minus 0.5ex}%
  {1sp}%
  {\normalfont\normalsize\centering}}
\makeatother

\def\BibTeX{{\rm B\kern-.05em{\sc i\kern-.025em b}\kern-.08em
    T\kern-.1667em\lower.7ex\hbox{E}\kern-.125emX}}

\newcommand{\ashish}[1]{{\color{red}{Ashish: #1}}}

\newcommand{\keiwan}[1]{{\color{magenta}{KS: #1}}}

\usepackage{fancyhdr}
\fancypagestyle{firstpage}{%
  \fancyhf{} 
  \fancyhead[C]{\footnotesize This paper has been accepted at the 22nd IEEE International Conference on Mobile Ad-Hoc and Smart Systems (MASS), 2025.}
}

\begin{document}

\title{Energy-Efficient Split Learning for Resource-Constrained Environments:\\
A Smart Farming Solution\\

}

\author{
	\IEEEauthorblockN{
	    	Keiwan Soltani\IEEEauthorrefmark{1}, Vishesh Kumar Tanwar\IEEEauthorrefmark{1}, Ashish Gupta\IEEEauthorrefmark{2}, Sajal K. Das\IEEEauthorrefmark{1}
	}
    \IEEEauthorblockA{
		\IEEEauthorrefmark{1}Department of Computer Science, Missouri University of Science and Technology, USA \\ \IEEEauthorrefmark{2}Department of Computer Science and Engineering, BITS Pilani Dubai Campus, UAE\\
        Email: $\left\{\text{ksoltani, vishesh.tanwar, sdas}\right\}$@mst.edu, ashish@dubai.bits-pilani.ac.in
	}
    \vspace{-0.2in}
}

\maketitle
\thispagestyle{firstpage}
\begin{abstract}
Smart farming systems encounter significant challenges, including limited resources, the need for data privacy, and poor connectivity in rural areas. To address these issues, we present \textit{eEnergy-Split}, an energy-efficient framework that utilizes split learning (SL) to enable collaborative model training without direct data sharing or heavy computation on edge devices. By distributing the model between edge devices and a central server, \textit{eEnergy-Split} reduces on-device energy usage by up to 86\% compared to federated learning (FL) while safeguarding data privacy. Moreover, SL improves classification accuracy by up to 6.2\% over FL on ResNet-18 and by more modest amounts on GoogleNet and MobileNetV2. We propose an optimal edge deployment algorithm and a UAV trajectory planning strategy that solves the Traveling Salesman Problem (TSP) exactly to minimize flight cost and extend and maximize communication rounds. Comprehensive evaluations on agricultural pest datasets reveal that \textit{eEnergy-Split} lowers UAV energy consumption compared to baseline methods and boosts overall accuracy by up to 17\%. Notably, the energy efficiency of SL is shown to be model-dependent—yielding substantial savings in lightweight models like MobileNet, while communication and memory overheads may reduce efficiency gains in deeper networks. These results highlight the potential of combining SL with energy-aware design to deliver a scalable, privacy-preserving solution for resource-constrained smart farming environments.
\end{abstract}




\section{Introduction}
By 2050, global agriculture must nearly double its output to meet the demands of an estimated 9.7 billion people~\cite{bhat2021big, ray2013yield}. 
While large farms are equipped to adopt advanced agriculture tools, small and medium-sized rural growers frequently lack the necessary infrastructure for the efficient implementation of available technologies (e.g., for pest detection) \cite{albanese2021automated,lima2020automatic}. Without timely interventions or accurate results, most farmers resort to excessive pesticide usage, degrading soil health, and contaminating water sources. Insights from our NSF I-Corps study~\cite{I-Corps2023} confirm the need for cost-effective, privacy-preserving technologies that remain functional in low-connectivity, resource-limited agricultural environments.

Unmanned aerial vehicles (UAVs) offer a promising solution to connectivity and data-processing limitations~\cite{lin2020energy,caruso2021collection,velusamy2021unmanned}. UAVs can operate as mobile sensing and computational platforms, bridging communication gaps and providing multi-sensory field data. By capturing localized pest activity and environmental conditions, UAVs reduce the reliance on centralized data aggregation, mitigating bandwidth constraints and latency issues. For instance, the authors in~\cite{jiang2022trust} proposed a trust-based scheme that ensures only reliable sensor data is collected, improving decision-making in pest management by filtering out faulty or malicious nodes. Moreover, the authors in~\cite{soltani2023trust} integrate trust mechanisms with energy-efficient data gathering in wireless sensor networks to enhance reliability and prolong network lifetime. While~\cite{xu2021minimizing,tyrovolas2022energy} focused on affordability and energy-aware UAV deployment methods, their findings about minimizing the number of UAVs and leveraging reconfigurable intelligent surfaces are equally relevant to optimizing large-scale agricultural operations.


Further, from an aspect of Machine Learning (ML) with privacy preserving concerns, Federated Learning (FL)~\cite{le2023federated} has been adopted recently, which enables collaborative global model training without direct data sharing. However, it can burden resource-constrained devices by requiring local training of the entire model, potentially exposing them to privacy attacks. For example, the work~\cite{Wu2022SplitLO} developed personalized FL frameworks to ensure that farms' conditions—soil moisture, pest incidence, and crop phenotypes—are adequately captured without overburdening weaker devices or networks. On the communication front, techniques like collaborative FL~\cite{Chen2020WirelessCF} to reduce reliance on a central controller, and optimized FL algorithms focusing on resource allocation~\cite{Nguyen2021EfficientFL}, demonstrate the network-aware learning strategies. 
Recent studies in UAV networks highlight the potential of integrating split learning (SL)~\cite{liu2022novel,thapa2022splitfed} with FL and reinforcement learning~\cite{tanwar2025reindsplit} to achieve enhanced scalability, accuracy, and energy efficiency. In SL, the model is partitioned between clients and a server to reduce clients' computational burden. Liu et al.~\cite{liu2022energy} presented a user scheduling approach that judiciously assigns participants to split or federated training modes to minimize energy consumption. 


\noindent $\bullet$ {\bf Motivation:}
This research is motivated by the limitations of prior works: (i) {\em data privacy} -- transmitting large volumes of raw data (collected by sensors deployed in farms) for remote model training is costly and raises privacy concerns; (ii) {\em UAV's energy constraints} -- as UAVs run on battery, finding an optimized tour for data/model collection from the edge devices (sensors with computational capabilities) is still a challenge; (iii) overburdening the edge devices with intensive training of a complete ML model.

In this paper, while addressing the above limitations, we solve the problem: {\em how to build an ML-based agriculture solution with resource-constrained IoT devices while preserving data privacy.} As a solution, we propose an energy-efficient split learning based framework (abbreviated as {\bf eEnergy-Split}).

\noindent $\bullet$ {\bf Major Contributions:}
\begin{itemize}
    \item We present a novel framework, eEnergy-Split, that facilitates model training by integrating a new method for optimal sensor deployment and an energy-aware UAV trajectory design using an exact Traveling Salesman Problem (TSP) solver, thereby minimizing the UAV's energy consumption and maximizing the number of completed communication rounds. 
    \item The eEnergy-Split framework strategically integrates SL with FL by dividing the model into two parts—one residing on edge devices and the other on the server—thereby reducing the computational load on edge devices during model training. 
    \item Through a series of insightful experiments, we evaluate the overall framework using benchmark models such as ResNet18, GoogleNet, and MobileNet V2, measuring metrics like energy consumption, CO$_2$ emissions, and accuracy. The results clearly show reduced energy usage, effectively extending network lifetime during incremental training.  
\end{itemize}

\noindent \textbf{Organization:} Section~\ref{system_model} discusses the system model and problem formulation. Section~\ref{proposed_methodology} introduces the \textit{eEnergy-Split} framework. We report the experimental results in Section~\ref{experiments}. Finally, Section~\ref{conclusion} concludes the paper with future research directions.

\section{System Model and Problem Description}\label{system_model}
This section first describes the system along with the adopted energy model and then mathematically formulates the problem to be solved.

\subsection{System Model}
This work considers a wireless sensor network of $N$ sensors $S=\{s_1, s_2, \dots, s_N\}$ uniformly distributed in an agricultural field $A$. Each sensor records data (e.g., soil moisture, humidity, amount of minerals, images of insects, etc.) from the monitoring environment. This work considers two types of sensors: (i) the ones that capture data from the environment and (ii) the ones (also referred to as {\em edge devices}) that also have computational capabilities to process the collected data. Let $E=\{e_1,e_2,\cdots, e_M\}$ be a set of M sensors of the latter type, where $E \subseteq S$ and $M \leq N$. 
Let $(x_i, y_i)$ denote the coordinates of an IoT sensor $s_i$, for $1 \leq i \leq N$. Let $0<\Delta_i\leq \Delta$ be the size of the recorded data by the sensor $s_i$. 
The distance (Euclidean) between $s_i$ and $s_j$ is denoted by $d_{s_i,s_j}$, and the sensors can communicate with each other within the Communication Range ($CR$) if $d_{s_i,s_j}\leq CR$.
Similarly, the UAV can successfully collect data from an edge device $e_i$ if $d_{UAV,e_i} \leq Rr$, where $Rr$ is the reception range of the UAV. We assume that if the UAV hovers on top of an edge device, it can establish an acceptable communication. The UAV's data reception range $Rr$ at altitude $h$ is determined as $ Rr=\sqrt{CR^2 - h^2}$~\cite{10723590}. Note that the communication challenges, such as multi-path propagation, fading, or shadowing, are out of the scope of this work.

\begin{figure*}[h]
    \centering
    \vspace{-0.12in}
    \includegraphics[width=\textwidth,height=5cm,keepaspectratio=false]{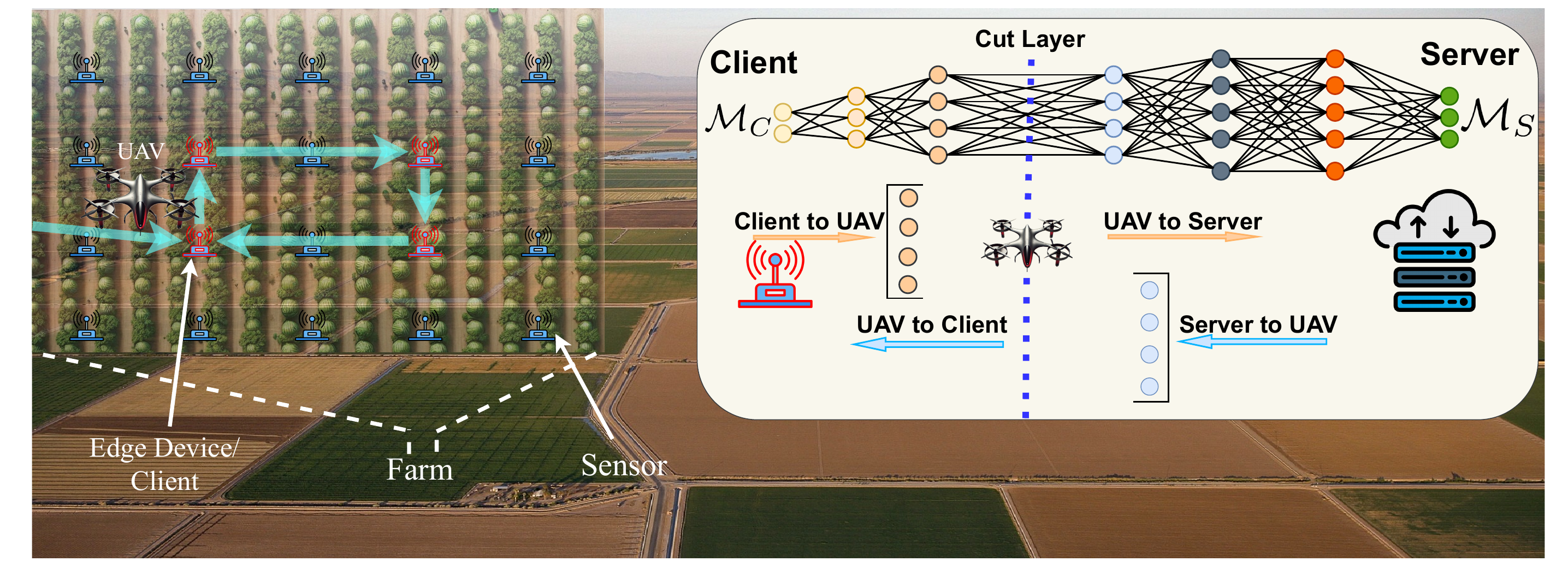}
    \caption{The left sub-figure depicts a single farm with an example of the UAV trajectory for exchanging data with the cluster head devices (red-colored sensors). The right sub-figure illustrates the communication between a cluster head (client/edge device), a UAV, and a server.}
    \vspace{-0.2in}
    \label{fig:UAV-Data-collection}
\end{figure*}







\noindent $\bullet$ {\bf Energy Model for UAV:}
The energy consumption of a UAV is determined by three primary components: 
(i) $\xi_m-$ energy required for movement between hover points, (ii) $\xi_h-$ the energy used to hover for data exchange, and (iii) $\xi_c-$ the energy consumed for communication purposes. 
Let $T_m, T_h,$ and $T_c$ respectively denote the total movement time, total hovering time, and total communication time (for data gathering, uploading, and downloading model weights), needed to complete a single tour, where $T=T_m+T_h+T_c$. 
Moreover, the total energy utilized by the UAV during the mission must be less than its energy budget $\beta$, i.e., $ T_m \cdot \xi_m + T_h \cdot \xi_h + T_c \cdot \xi_c< \beta$, where energy consumption components are computed for rotary-wing UAV~\cite{8663615} using Eq.~\eqref{eq:eng_movement} and Eq.~\eqref{eq:eng_hovering}.

\vspace{-0.1in}
\begin{small}
\begin{equation}\label{eq:eng_movement}
    \xi_m = P_0 (1+ \frac{3V^2}{U_{tip}^2}) + P_i (\sqrt{1+ \frac{V^4}{4v_0^4}} - \frac{V^2}{2v_0^2})^{1/2} + \frac{1}{2}f\rho r aV^3,
\end{equation}
\begin{equation}\label{eq:eng_hovering}
    \xi_h = P_0 + P_i,
\end{equation}
\end{small}
where
 $P_0 = \frac{\delta}{8}\rho r a \Omega^3 R^3  \text{ and } P_i = (1+ k) \frac{W^{3/2}}{\sqrt{2\rho a}}$, and
\noindent $P_0$ and $P_i$ represent the blade profile power and induced power, respectively.
Table~\ref{tbl:UAV-parameters} shows the meaning and values for the other parameters that are selected based on the DJI Matrice 350 RTK, a widely used industrial-grade UAV known for its high endurance, stability, and advanced flight capabilities~\cite{DJI_M350}. 
The reason for using the weight $W$ in Newtons is to accurately represent the gravitational force acting on the UAV, as it is directly relevant to the power required for hovering and propulsion, which depend on force rather than mass. The weight $W$ can be calculated in Newtons using $W = m \cdot g$, where $m$ is the mass in kilograms, and $g$ is the acceleration due to gravity (9.81 $m/s^2$).  

\vspace{-0.15in}
\begin{table}[h]
\scriptsize
\caption{The parameters of the UAV energy model.}
\vspace{-0.15in}
\begin{center}
\renewcommand{\arraystretch}{1.2}
\setlength{\tabcolsep}{4pt}
\begin{tabular}{c|p{0.45\columnwidth}|c}
\cline{1-3} 
\textbf{\textit{Parameters}}& \textbf{\textit{Meaning}} & \textbf{\textit{Value}} \\
\hline
$\beta$&  UAV energy capacity  &   1.9 $MJ$\\
\hline
$V$&  UAV speed &   10 $m/s$\\
\hline 
$v_0$& Mean motor induced velocity in hover&     5.5 $m/s$\\	
\hline
$U_{tip}$&  Tip speed of the rotor blade&    180 $m/s$\\
\hline 
$f$&    Fuselage drag ratio&  0.8\\
\hline 
$r$&    Rotor solidity&   0.08\\
\hline 
$\rho$&    Air density&   1.225 $kg/m^3$\\
\hline 
$a$&    Rotor disc area&   0.7 $m^2$\\
\hline 
$\delta$&   Profile drag coefficient&   0.011\\
\hline 
$\Omega$&   Blade angular velocity&   320 $radians/s$\\
\hline 
$R$&   Rotor radius&   0.45 $m$\\
\hline 
$k$&    Incremental correction factor to induce power&  0.15\\
\hline 
$W$&    Aircraft weight&  63.4 $Newton$\\
\hline 
\end{tabular}
\label{tbl:UAV-parameters}
\end{center}
\vspace{-0.25in}
\end{table}

\noindent $\bullet$ {\bf Split Learning for Model Training:}
Training deep learning models is computationally intensive, which can quickly drain the batteries of edge devices. Conversely, training on the cloud requires uploading all data, often necessitating numerous UAV tours that consume significant energy and shorten network lifetime. Therefore, reducing the computational burden of model training is essential, and SL offers a promising solution. In SL, training is divided: a portion is handled on the client side, while the remainder is done by a powerful cloud server.
This paper considers a scenario where sensors periodically capture data from an agricultural field, and edge devices perform the client-side training. Due to communication constraints, these devices cannot directly forward their client-side smashed data or receive updated weights from the central server. Consequently, a UAV is employed to collect data from the edge devices, upload it to the cloud central server, and download the updated weights for broadcasting among the edge devices. The UAV facilitates this by traveling between predefined locations obtained strategically in the next section.
\vspace{-0.05in}
\subsection{Problem Description}


This research aims to jointly determine: (i) optimal positions for deploying edge devices to maximize data collection from the sensors and (ii) an efficient UAV trajectory that minimizes both the number of edge devices $|E|$ and the total UAV travel distance $D_{UAV}$, while maximizing the number of complete UAV tours, $\gamma$, over edge devices. The objective function as expressed in Eq.~\eqref{eq:objective} considers normalized values of the concerned parameters.
Eq.~\eqref{eq:coverage} enforces that the aggregation of the coverage of each edge device ($e_i$) captured as $C(e_i)$ covers entire set of the sensors deployed ($S$). 
At the same time, the UAV's energy constraint $\beta$ as expressed in Eq~\eqref{eq:energy} must be respected. In SL, the additional communication overhead for exchanging the raw and updated smashed data is denoted by $T_{SL}$.
Formally, the optimization problem is defined as:

\vspace{-0.1in}
\begin{align}
    \min_{M,\, D_{UAV},\, \gamma} & \quad M + D_{UAV} - \gamma, \label{eq:objective}\\[1mm]
    \text{Subject to:}\hspace{0.05cm} & \bigcup_{e_i \in E} \mathcal{C}(e_i) = S, \label{eq:coverage}\\
    & \xi_m+\xi_h+\xi_c < \beta, \label{eq:energy}\\
    & \gamma = \left\lfloor \frac{\beta}{\, T_m\cdot \xi_m + T_h\cdot \xi_h + T_c\cdot \xi_c } \right\rfloor, \label{eq:tour}\\
    & D_{UAV} = \sum_{i=1}^{M} d(e_i, e_{i+1}), \label{eq:distance}\\
    & T_{SL} = \frac{L}{R} \label{eq:T_SL}
\end{align}
In Eq.~\eqref{eq:tour}, the term
$T_m\cdot \xi_m + T_h\cdot \xi_h + T_c\cdot \xi_c$ quantifies the total energy expenditure. Eq.~\eqref{eq:distance} expresses the total travel distance for a UAV.
In Eq.~\eqref{eq:T_SL}, $L$ denotes the size of the smashed data generated by the partial model at the client, and $R$ represents the effective data rate for UAV-edge communications.


\section{\MakeLowercase{e}Energy-Split Framework}\label{proposed_methodology}
In this section, we propose an energy-efficient split learning framework ({\bf eEnergy-Split}) comprising three major strategies for: (i) sensors and edge devices deployment, (ii) UAV trajectory design, (iii) SL based model training on edge devices.
eEnergy-Split framework considers a UAV-assisted SL for precision agriculture scenarios, especially pest detection, with constrained connectivity and limited on-device resources. The overview of the framework is shown in Fig.~\ref{fig:UAV-Data-collection}, in which sensors collect data and hand over it to an edge device, on which a partial deep learning model $\mathcal{M}_C$ (with the first few layers) is trained. Later, the ground truth,
along with the client-side model weights, are offloaded to a UAV (visits periodically) that further delivers the received load to a server that hosts the remaining portion $\mathcal{M}_S$. The UAV operates as a mobile data relay between edge devices and the server. This paradigm reduces the computational burden on resource-limited devices and preserves data privacy while never exposing raw data.

\vspace{-0.05in}
\subsection{Edge Devices Deployment}
This section introduces an algorithm to deploy sensors (with limited hardware) and edge devices (sensors with computational resources like Jetson Nano) to optimize the overall data collection and model training process. 
Each edge device serves as a representative for a group of normal sensors and receives the collected data (image or other captured information from the field) from the sensors. To efficiently select and deploy edge devices, we propose an Algorithm~\ref{alg:jetson_deployment}. 


\begin{algorithm}[t]
\small
\caption{Optimized Edge Device Deployment and Sensor Assignment using CSR}
\label{alg:jetson_deployment}
\begin{algorithmic}[1]
\Require Sensors $S = \{s_1,\dots, s_N \}$, communication\_range $CR$
\Ensure Edge device positions and sensor assignments

\State Construct adjacency list $\mathcal{A}$ with pair wise distances
\State $\mathcal{A}[s] \gets \{u \mid d_{s,u} \leq CR, u \in S\}$ \Comment{stored in CSR format}
\State $U \gets \{s_1, s_2, \dots, s_N\}$ \Comment{Uncovered sensors}
\State $E \gets \emptyset$ \Comment{Edge device positions}

\While{$U \neq \emptyset$}
    \State $\widehat{\Delta} \gets 0$ \Comment{Maximum coverage}
    \State $\widehat{s} \gets \text{None}$ \Comment{Best sensor for edge device}
    \For{each $s \in U$}
        \State $\widehat{C} \gets \mathcal{A}[s] \cap U$ \Comment{Covered sensors by $s$}
        \If{$|\widehat{C}| > \widehat{\Delta}$ \textbf{and} $E = \emptyset$ }
            \State $\widehat{\Delta} \gets |\widehat{C}|$
            \State $\widehat{s} \gets s$
        \ElsIf{$|\widehat{C}| \geq \widehat{\Delta}$ \textbf{and} $E \neq \emptyset$ \textbf{and} $\sum d_{s,E} < \sum d_{\widehat{s}, E}$}
            \State $\widehat{\Delta} \gets |\widehat{C}|$
            \State $\widehat{s} \gets s$
        \EndIf
    \EndFor
    \State $E \gets E \cup \{\widehat{s}\}$ \Comment{Place Edge device at $\widehat{s}$}
    \State $U \gets U \setminus \widehat{C}$ \Comment{Update uncovered sensors}
\EndWhile

\hspace{-0.3in}\textbf{Assign Sensors to Edge Devices}
\For{each $s \in S \setminus E$}
    \State Identify candidate edge devices within $CR$ from CSR adjacency list $\mathcal{A}$
    \State Select edge device with minimal current load and shortest distance
    \State Assign sensor $s$ to this selected edge device
    \State Update load distribution of edge devices
\EndFor

\State \Return Edge devices $E$ and sensor assignments
\end{algorithmic}
\end{algorithm}

In Algorithm~\ref{alg:jetson_deployment}, the deployment is achieved by prioritizing sensor nodes that provide the maximum coverage of their neighboring sensors within the communication range ($CR$). Using compressed sparse row (CSR) representation, the algorithm efficiently maintains adjacency information, significantly reducing memory usage and computational overhead. From the set of uncovered sensors, the sensor with the highest coverage degree is initially selected as an edge device. In the case of a tie, where multiple sensors cover the same number of uncovered nodes, the selection process further optimizes placement by choosing the sensor that minimizes the total distance to the already placed edge devices. This ensures an efficient distribution of edge devices across the network. Once an edge device is placed, all the uncovered sensors in its adjacency are removed from the uncovered set. The process continues iteratively until all sensors in the network are covered. The selected sensors as edge devices are the designated spots for client-side computation and UAVs' optimal hovering locations.

After the deployment of edge devices, the next crucial step is the assignment of sensors to their respective edge devices. This assignment must be optimized to balance the computational load and maintain efficient communication. Therefore, from Lines 22 to 27 of Algorithm~\ref{alg:jetson_deployment}, we introduce an optimized sensor-to-edge device assignment strategy.



\noindent{\bf Time Complexity:} Algorithm~\ref{alg:jetson_deployment} iteratively examines all uncovered sensors, leading to a worst-case complexity of $O(N^2)$, where $N$ is the number of sensors.
The construction of the CSR-based adjacency list involves computing pairwise distances with a complexity of $O(N^2)$, which dominates the preprocessing phase.

\vspace{-0.1cm}
\subsection{Energy-Aware UAV Trajectory Design}

\begin{algorithm}[h]

\small
\caption{Energy-Constrained UAV Tour Planning Using Exact TSP Solver}
\label{alg:uav_optimized_path}
\begin{algorithmic}[1]
\Require UAV's initial position $O$, edge device positions $E = \{e_1, e_2, \dots, e_M\}$, energy budget $\beta$, UAV speed $V$, movement energy $\xi_m$, hovering energy $\xi_h$, communication energy $\xi_c$
\Ensure Optimal UAV tour $\pi$, number of full rounds $\gamma$

\State Construct a complete distance graph $G$ where nodes are $E$, edges are Euclidean distances between all device pairs
\State Solve the exact TSP on $G$ to find minimal tour $\pi$ visiting all nodes in $E$
\State Append the first node $e_1$ to the end of $\pi$ to complete the tour

\State \textbf{Energy and Round Initialization:}
\State $D_\pi \gets$ Total tour length of $\pi$
\State $\widehat{E}_\pi \gets \frac{D_\pi \cdot \xi_m}{V} + M \cdot (\xi_h + \xi_c)$ \Comment{Per round energy: move + hover + comm}
\State $e_1 \gets \pi[0]$, $e_M \gets \pi[-2]$ \Comment{Start and last edge device in tour}
\State $\widehat{E}_\text{first} \gets \frac{d_{O, e_1} \cdot \xi_m}{V} + \widehat{E}_\pi$
\State $\widehat{E}_\text{return} \gets \frac{d_{e_M, O} \cdot \xi_m}{V}$ \Comment{Return energy}
\State $\gamma \gets 0$, $\widehat{\beta} \gets \beta$

\If{$\widehat{E}_\text{first} + \widehat{E}_\text{return} > \widehat{\beta}$}
    \State \Return $\pi$, $0$ \Comment{Insufficient energy for one full round}
\EndIf

\State $\widehat{\beta} \gets \widehat{\beta} - \widehat{E}_\text{first}$
\State $\gamma \gets 1$

\While{$\widehat{\beta} \geq \widehat{E}_\pi + \widehat{E}_\text{return}$}
    \State $\widehat{\beta} \gets \widehat{\beta} - \widehat{E}_\pi$
    \State $\gamma \gets \gamma + 1$
\EndWhile

\State \Return $\pi$, $\gamma$
\end{algorithmic}

\end{algorithm}

This section presents an energy-aware UAV trajectory planning algorithm that determines the most efficient route for the UAV (i.e., path through the edge devices). After deployment, each sensor transmits its data to the assigned edge devices periodically. Once an edge device has received all data from its associated sensors, it performs client-side computation, i.e., partial model training. 
The computed model weights from edge devices, unable to directly reach the cloud server, are collected and relayed by a UAV for central training and aggregation. Subsequently, the UAV distributes the updated weights back to the devices for the next training iteration. To support this process, the UAV must follow an energy-efficient route to maximize the number of complete training rounds before needing to recharge.

Since UAVs are energy-constrained, optimizing their flight path is essential to maximizing the number of communication rounds they can perform before battery replacement or recharging is required. Because the number of edge devices is relatively small, we frame the UAV routing task as a classical TSP, aiming to find the shortest possible tour that visits all edge devices exactly once. Unlike heuristic-based solutions, we adopt an exact TSP solver that guarantees the globally optimal tour, minimizing the total flight distance.
Although exact TSP has exponential worst-case complexity, our deployments involve only a few edge devices for farms up to 250 acres, enabling optimal routes to be computed almost instantly. For larger-scale scenarios, the method can be adapted to use heuristics to maintain near-optimal performance with practical runtimes.

The generated optimal tour $\pi$ does not include the base station as a mandatory node. Instead, the UAV departs from the base station to the first edge device, traverses the TSP tour, and repeats the tour as many times as energy ($\beta$) permits without returning to base after each round. Before initiating a new round, the UAV checks whether it has enough remaining energy to complete the next tour and return safely to the base station. If not, it aborts the next round and returns. This delayed-return strategy is incorporated in Algorithm~\ref{alg:uav_optimized_path}, which captures the full energy-aware tour planning logic.

\begin{algorithm}[h]
\small
\caption{SL training in edge device-UAV framework}
\label{alg:custom_split_learning}
\begin{algorithmic}[1]
\Require
Devices $E = \{e_1, \dots, e_M\}$ with model $\mathcal{M}_C$ and dataset $\{D_{e} \in E\}$. Server with model $\mathcal{M}_S$, loss function $\mathcal{L}$, global aggregation rounds $R$, local split rounds $r$
\Ensure Updated $\mathcal{M}_S$ and $\mathcal{M}_C$ performance, consume energy

\State Initialize parameters for $\mathcal{M}_C$ and $\mathcal{M}_S$
\State Initialize energy/time accumulators $(\mathcal{E}_{total}, T_{total})$

\For{$\text{global\_round} = 1$ \textbf{to} $R$}
    \For{each client $e \in E$ \text{and} $r$ rounds}
        \State Obtain mini-batch $(X, Y)$ from $D_e$
        \State \textbf{// Client Forward Pass:} 
        \State $(\mathcal{E}_{e}^{fwd},T_{e}^{fwd})$ $ \leftarrow$ EnergyTracker
        \State Compute smashed data
        \State \textbf{Record $T_{e}^{fwd}$ and $\mathcal{E}_{e}^{fwd}$}
        
        \State \textbf{// Server Forward Pass:}
        \State Compute server output and loss
        
        
        \State \textbf{// Distributed Backward Pass:}
        \State $(\mathcal{E}_{e}^{bwd}, T_{e}^{bwd}) $ $ \leftarrow$ EnergyTracker
        \State Backpropagate $\ell$ through $\mathcal{M}_S$ and $\mathcal{M}_C$
        \State Update parameters in $\mathcal{M}_S$ and $\mathcal{M}_C$

        \State $\mathcal{E}_{total} \gets \mathcal{E}_{total} + (\mathcal{E}_{e}^{fwd} + \mathcal{E}_{e}^{bwd})$
        \State $T_{total} \gets T_{total} + (T_{e}^{fwd} + T_{e}^{bwd})$
    \EndFor
    \State \textbf{// Device's Model Aggregation across $E$:}
    \[
        \theta_{agg} \gets \frac{1}{M} \sum_{e \in E} \theta_e
    \]
    \State Update each client $e$ with $\theta_{agg}$ 
\EndFor
\State \Return Final model $\mathcal{M}$ and $\mathcal{E}_{total}, T_{total}$
\end{algorithmic}
\end{algorithm}
\begin{figure*}[!t]
    \centering
    \vspace{-0.25in}
    \begin{subfigure}[t]{0.32\textwidth}
        \includegraphics[width=\linewidth]{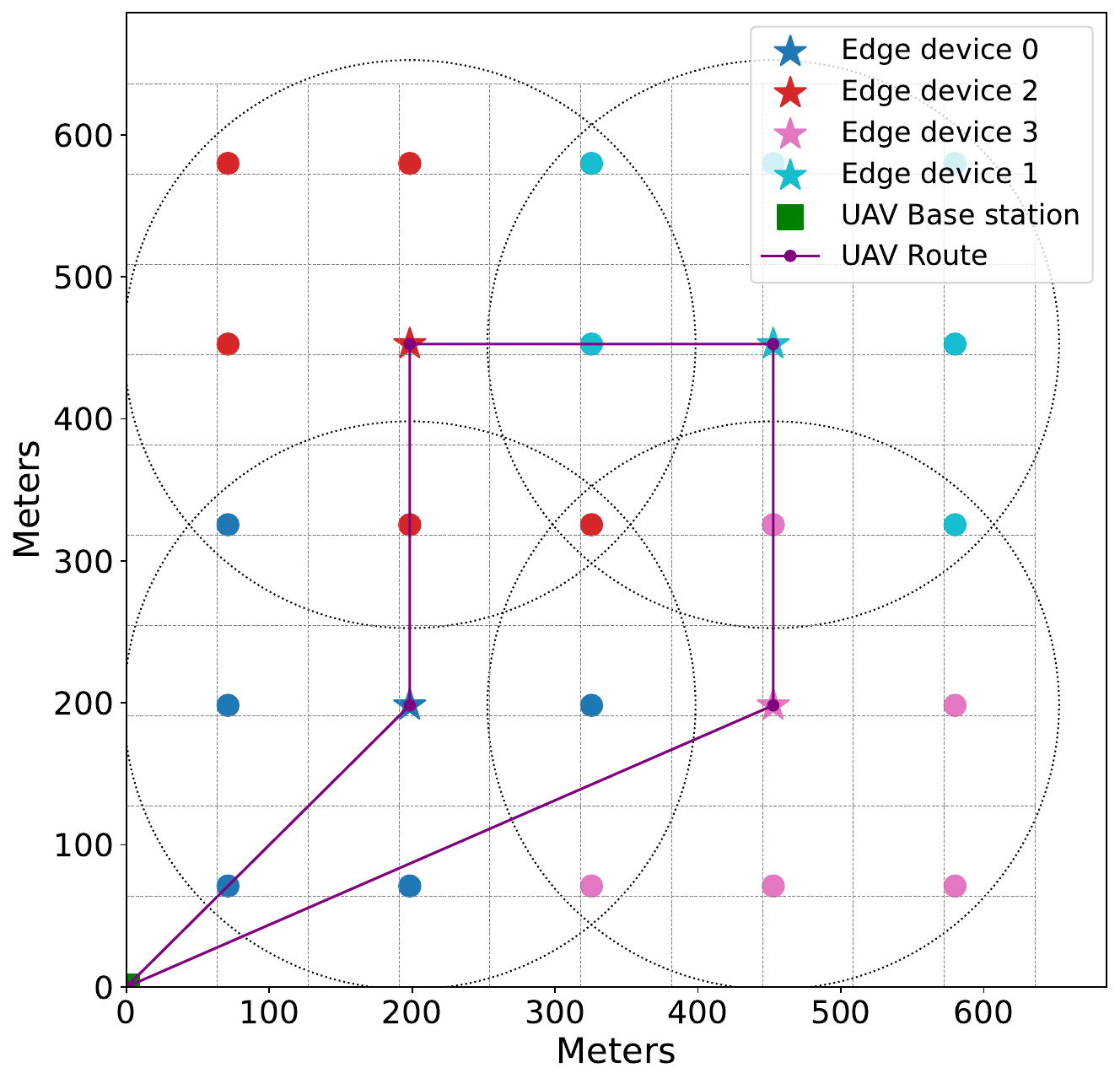}
        
        \caption{Our proposed method (eEnergy-Split)}
        \label{fig:proposed_method}
    \end{subfigure}
    \hfill
    \begin{subfigure}[t]{0.32\textwidth}
        \includegraphics[width=\linewidth]{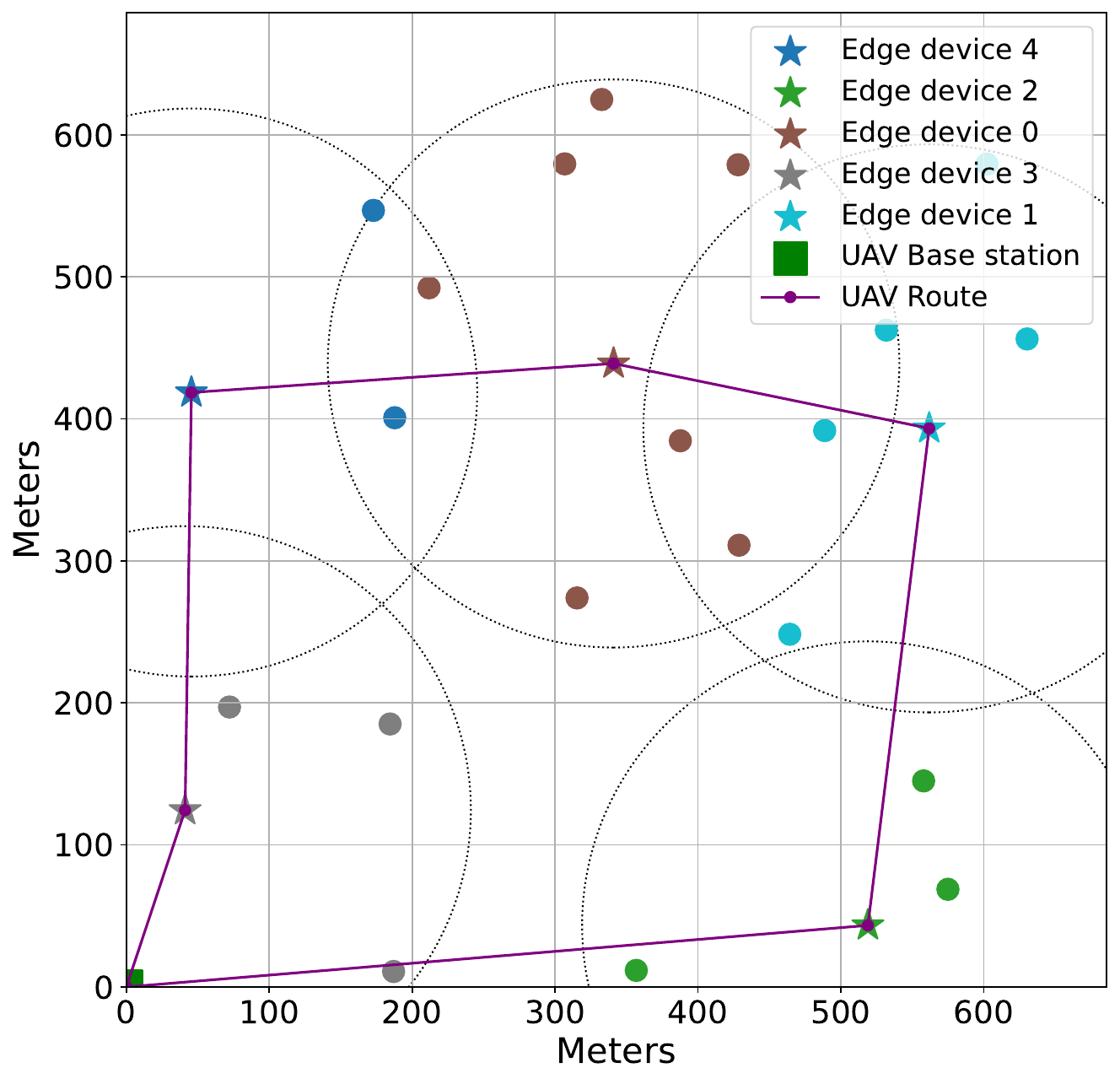}
        \caption{GASBAC}
        \label{fig:GASBAC}
    \end{subfigure}
    \hfill
    \begin{subfigure}[t]{0.32\textwidth}
        \includegraphics[width=\linewidth]{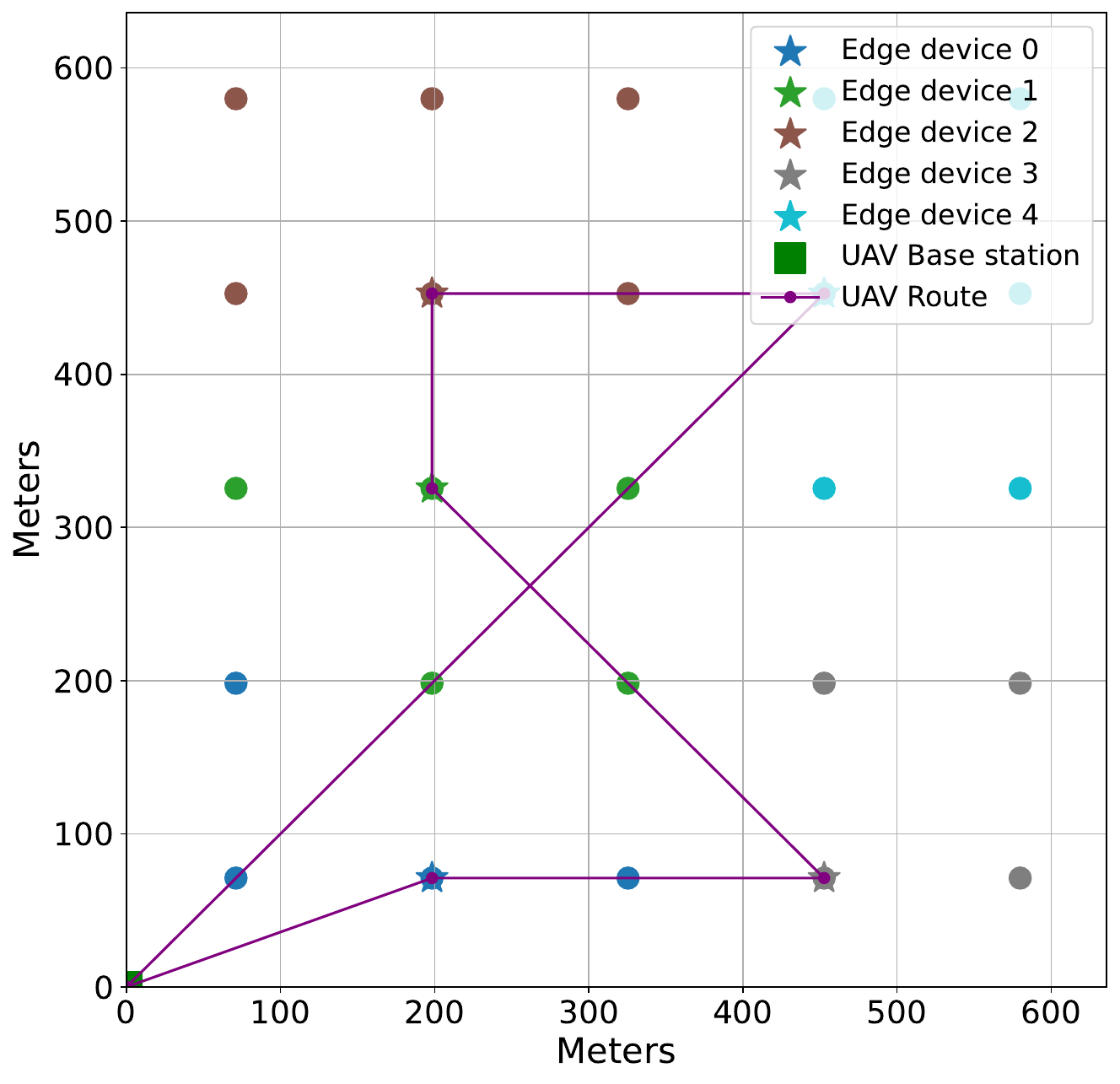}
        \caption{K-means}
        \label{fig:K-means}
    \end{subfigure}
    \caption{Comparison of the three deployment strategies.}
    \vspace{-0.2in}
    \label{fig:deplyment-strategies}
\end{figure*}
\vspace{-0.02in}
\subsection{SL Training under eEnergy-Split Framework}
Finally, the collaborative model training proceeds in alternating client ($\mathcal{C}$) and server ($\mathcal{S}$) side phases that the UAV orchestrates, as reported in Algorithm~\ref{alg:custom_split_learning}. At the beginning of each global round $R$, each edge device $e\in E$ holds a copy of the current client sub-model $\mathcal{M}_C$ and a local mini-dataset $D_e$. The edge device selects a mini-batch $(X, Y)\subset D_e$, forwards the raw samples through $\mathcal{M}_C$, and produces the smashed representation $Z$ to be handed over once the UAV arrives. The EnergyTracker routine records the time $T^{\text{fwd}}_{e}$ and the energy $\mathcal{E}^{T_{e}^{fwd}}$ consumed by the device during computation and short-range transmission.

The UAV relays $Z$ with the associated ground-truth labels to the server, where the remaining layers $\mathcal{M}_S$ generate the predictions and the server performs its portion of back-propagation, returning the gradient concerning $Z$. Upon receipt, the edge device continues the backward pass through $\mathcal{M}_C$; EnergyTracker again, measuring the client-side time $T^{\text{bwd}}_{e}$ and the energy $\mathcal{E}^{T_{e}^{bwd}}$. Both sub-models then update their parameters using local optimizers, ensuring that computation remains split yet synchronous. Once each client has completed $r$ such split rounds, the server aggregates the updated full-model ($\mathcal{M}$) weights with the arithmetic mean, $\theta_{agg}$, and the UAV will broadcast to all devices during its next flight. The accumulators $\mathcal{E}_{\text{total}}$ and $T_{\text{total}}$ thus capture the complete resource footprint of forward computation, backward computation, and wireless transfer across the network. After the final round, the server holds the converged models $(\mathcal{M}_C, \mathcal{M}_S)$.

\section{Experimental Evaluation}\label{experiments}

We evaluate the performance of our proposed algorithms in terms of sensor and edge devices deployment, energy consumption, and load distribution.

\subsection{UAV Trajectories}
Considering 25 sensors, with a communication range of 200 meters, deployed in a 100-acre farm, Fig.~\ref{fig:deplyment-strategies} demonstrates the deployment outcomes for eEnergy-Split (essentially Algorithm~\ref{alg:jetson_deployment}) and two baselines: Gathering data Assisted by Multi-UAV with a BAlanced Clustering (GASBAC)~\cite{GASBAC2023} and K-means. GASBAC is a set of heuristic algorithms that balance energy consumption across sensor clusters and optimize UAV trajectories to extend the operational lifetime of wireless sensor networks in precision farming. In K-means, the value of $K$ is initially set to $\lfloor \sqrt{N} \rfloor$ and incremented if any sensors remain unassigned. Once the edge devices and their corresponding sensor clusters are determined, the UAV follows a greedy approach to visit the edge devices.
In Figs.~\ref{fig:proposed_method} and~\ref{fig:K-means}, sensors are uniformly deployed at a density of one sensor per five acres, while in Fig.~\ref{fig:GASBAC}, sensors are randomly deployed. 
As shown in Fig.~\ref{fig:proposed_method}, based on the communication range ($CR$), all sensors' data can be collected using the minimal number of edge devices while also minimizing the distances between them. Additionally, sensors are assigned to edge devices in a way that ensures an optimal workload distribution, enhancing overall efficiency.


\begin{figure*}[!ht]
			\centering	
            \vspace{-0.15in}
            \includegraphics[scale=0.25]{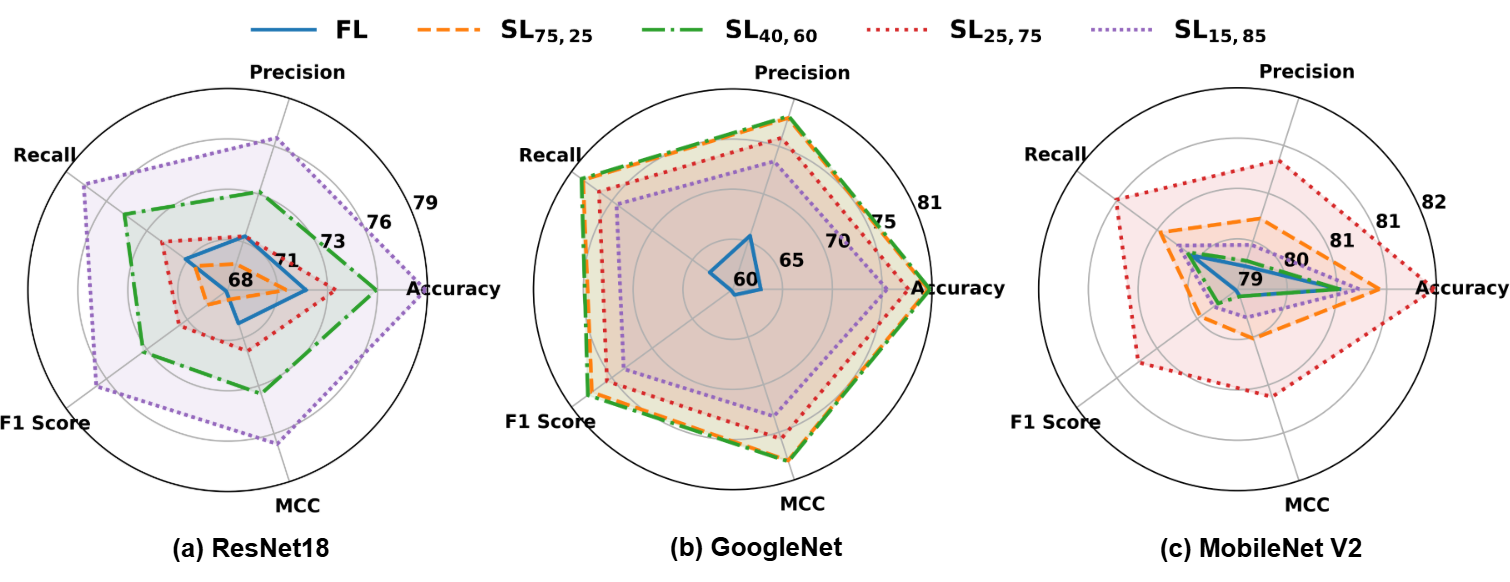}
			\vspace{-0.1in}
			\caption{Radar charts comparing evaluation metrics on the KAP dataset for different architectures and SL variations. Higher values closer to the outer edge indicate better performance. Note: SL$_{a, b}$ denotes that the client has a\% of layers and the server has the remaining b\% of layers.}
            \vspace{-0.2in}
		\label{evaluation_metrics} 
\end{figure*}

\subsection{UAV Energy Consumption}
Next, we analyze the energy consumption of the UAV for a complete tour under different deployment scenarios. As shown in Table~\ref{tbl:energy_consumption}, three distinct configurations are evaluated by varying the size of the agricultural field, the number of sensors, and the sensor-to-acreage ratio. 


Across all configurations, eEnergy-Split framework consistently outperforms K-means and GASBAC in terms of UAV energy efficiency. Specifically, in the 100-acre scenario, the UAV consumes only 35.1 kilojoules (kJ) using our approach, compared to 80.89 kJ and 92.80 kJ for K-means and GASBAC, respectively. This corresponds to a reduction of approximately $56.6\%$ and $62.2\%$ in energy consumption. Similarly, in the 140-acre field with 36 sensors, our method reduces energy use to 57.68 kJ, which is about $50\%$ and $51\%$ lower than K-means and GASBAC, respectively. In the most demanding 200-acre scenario with 49 sensors, the UAV energy consumption using our method is 103.10 kJ, significantly less than the 154.19 kJ for K-means and 164.37 kJ for GASBAC.

\begin{table}[h]
\centering
\caption{Energy in Joules (J) Consumption Analysis}
\renewcommand{\arraystretch}{1.4} 
\setlength{\tabcolsep}{4pt}       
\resizebox{\columnwidth}{!}{ 
\begin{tabular}{|c|c|c|c|c|c|}
\hline
\multicolumn{3}{|c|}{\textbf{Configuration}} & \multicolumn{3}{|c|}{\textbf{UAV Energy Consumption (kJ/Trip)}} \\ \hline
\textbf{Farm Acres} & \textbf{\# Sensors}& \textbf{Sensor per \# Acres} & \textbf{eEnergy-Split}& \textbf{K-means}& \textbf{GASBAC}\\ \hline
100& 25& 5& 35.07& 80.89& 92.80\\ \hline
140& 36& 4& 57.68& 114.96& 117.33\\ \hline
200& 49& 8& 103.10& 154.19& 164.37\\ \hline
\end{tabular}
}
\vspace{-0.1in}
\label{tbl:energy_consumption}
\end{table}

On average, our proposed method decreases UAV energy consumption by approximately $50\%$ compared to K-means and by $60\%$ compared to GASBAC. These improvements can be attributed to optimized edge device placement and sensor assignment strategies used in eEnergy-Split, which result in shorter UAV flight paths and more efficient data aggregation. In contrast, GASBAC, originally designed for multi-UAV systems, incurs higher overhead when adapted to a single UAV, while K-means suffers from suboptimal cluster formation. Overall, eEnergy-Split gives the most energy-efficient and scalable solution among the evaluated approaches.





\subsection{Performance Analysis}\label{perf_ana}
We implement all experiments in \textit{Python 3} using PyTorch, utilizing an Nvidia RTX A5000 GPU, 8 CPU cores $@$1.5 GHz (FP32$\approx$27.8 TFLOPS), 768 GB/s, tensor FLOPS$\approx$216 TFLOPS, CPU PassMark$\approx$35000, and 40 GB of system memory. We performed our experiments on three DNN architectures: ResNet18 (RN)~\cite{he2016deep}, GoogleNet (GN)~\cite{szegedy2015going}, and MobileNetV2 (MN)~\cite{sandler2018mobilenetv2} on Kaggle's Agriculture Pests (KAP) image dataset~\cite{kap}, having 12 pest species, namely, ants, bees, beetles, caterpillars, moths, earthworms, earwigs, grasshoppers, slugs, snails, wasps, and weevils. We partition the training dataset into training (90\%) and validation (10\%), and report the results of the test dataset, and images are resized to \texttt{(224, 224)}. We split each model into four sub-models to simulate the SL in low-configured edge devices (with 15\%, 25\%, 40\%, and 75\% of the layers, and the remainder is over the server) and compare client and server computational loads with different cut layers.\\\vspace{-0.25cm}

\noindent \textbf{Hyperparameter and Data Heterogeneity:} We consider four clients and assign data of 3 classes to each to simulate a non-IID distribution. We use the cross-entropy loss function, the \texttt{AdamW} optimizer, and evaluate on metrics such as accuracy, precision, recall, F1-score, and Matthews Correlation Coefficient (MCC). Fig.~\ref{evaluation_metrics} compares the classification performance of FL and SL across three backbone architectures: ResNet‑18 (a), GoogleNet (b), and MobileNetV2 (c).  For SL, we vary the fraction of layers hosted on the client from 75\% down to 15\% (i.e., SL$_{75,25}$ in orange dashed, SL$_{40,60}$ in green dash-dot, SL$_{25,75}$ in red dotted, and SL$_{15,85}$ in purple dotted). 

For \textbf{ResNet-18}, the FL baseline attains 72.34\% accuracy, F1 = 0.684, and MCC = 0.701.  Leveraging SL yields steady gains as more layers move server-side: the orange dashed trace (SL$_{75,25}$) shows 71.34\% accuracy (–1.0 percentage point (pp) vs. FL), F1 = 0.697, and MCC = 0.688; the green dash-dot curve (SL$_{40,60}$) rises to 75.98\% (+3.6 pp), F1 = 0.736, and MCC = 0.739; the red dotted line (SL$_{25,75}$) gives 73.89\% (+1.6 pp), F1 = 0.712, and MCC = 0.716; and the purple dotted trace (SL$_{15,85}$) achieves the best performance with 78.53\% (+6.2 pp), F1 = 0.765, and MCC = 0.766.  Thus, SL$_{15,85}$(purple) exceeds FL by over 6\% in accuracy and at least 8\% F1.

For \textbf{GoogleNet}, the FL baseline delivers only 63.15\% accuracy, F1 = 0.605, and MCC = 0.609, whereas the red dotted trace (SL$_{25,75}$) jumps to 80.35\% (+17.2 pp), F1 = 0.785, and MCC = 0.786.  Both the purple dotted (SL$_{15,85}$, 80.16\% Acc, F1 = 0.780, MCC = 0.785) and green dash-dot (SL$_{40,60}$, 78.16\% Acc, F1 = 0.761, MCC = 0.762) remain above FL, confirming that GoogleNet’s inverted-residual design is highly amenable to SL.

In the case of \textbf{MobileNetV2}, we begin from a strong FL baseline of 80.62\% accuracy, F1 = 0.788, and MCC = 0.789.  SL$_{75,25}$ further improves to 81.35\% (+0.73 pp), F1 = 0.796, and MCC = 0.797. SL$_{40,60}$ (green dash-dot line) holds steady at 80.62\% accuracy, F1 = 0.792, and MCC = 0.789, whereas SL$_{25,75}$ achieves the peak performance with 82.35\% accuracy (+1.73 pp), F1 = 0.810, and MCC = 0.808.  Finally, SL$_{15,85}$ delivers 80.98\% accuracy (+0.36 pp), F1 = 0.793, and MCC = 0.793, nearly matching the FL baseline.

These results demonstrate that SL, when the server hosts at least 60\% of layers, can consistently match or outperform FL on all three architectures.  Moreover, the optimum split point varies by backbone (most server-heavy for ResNet-18 and GoogleNet, intermediate for MobileNetV2), illustrating how SL enables a fine-grained trade-off between client computing, privacy exposure, and end-to-end accuracy.

\vspace{-0.1in}
\subsection{Resource Efficiency Analysis}
We estimate the execution time on a low-powered edge device (Jetson AGX Orin~\cite{jetson_orin}) by scaling our measured times on an NVIDIA RTX A5000 according to key hardware metrics. Specifically, we account for differences in FP32 throughput (FLOPS), memory bandwidth ($\mathrm{MemBW}$), tensor-core performance ($\mathrm{TFLOPS}$), CPU capability, software factor ($\mathrm{SF})$, and optimization factor ($\mathrm{OF})$. We measured the source device timing $T_{\mathrm{src}}$ on RTX A5000 and estimated timing $T_{\mathrm{tgt}}$ on Jetson as: 
\begin{equation}
\label{eq:conversion_formula}
\begin{aligned}
T_{\mathrm{tgt}}
=\,T_{\mathrm{src}}
&\times\Bigl(\tfrac{\mathrm{FLOPS}_{\mathrm{src}}}
                    {\mathrm{FLOPS}_{\mathrm{tgt}}}\Bigr)^{w_1} \times\Bigl(\tfrac{\mathrm{MemBW}_{\mathrm{src}}}
                    {\mathrm{MemBW}_{\mathrm{tgt}}}\Bigr)^{w_2}\\
&\times\Bigl(\tfrac{\mathrm{TFLOPS}_{\mathrm{src}}}
                    {\mathrm{TFLOPS}_{\mathrm{tgt}}}\Bigr)^{w_3}
\times\Bigl(\tfrac{\mathrm{CPU}_{\mathrm{src}}}
                    {\mathrm{CPU}_{\mathrm{tgt}}}\Bigr)^{w_4}\\
&\times \mathrm{SF} \times \mathrm{OF},
\end{aligned}
\end{equation}

\noindent where
$w_1=1.0,\;w_2=0.5,\;w_3=0.8,\;w_4=0.3, \mathrm{SF}=\mathrm{OF}=1.$
For simulation, we consider Jetson AGX Orin having 2048 CUDA cores, $@$1.3 GHz (FP32$\approx$2.7 TFLOPS), 51.2 GB/s, tensor FLOPS$\approx$21.6 TFLOPS, and CPU PassMark$\approx$2500. We assumed that the time on RTX A5000 was measured under identical software and batch sizes. Energy and CO$_2$ emissions on Jetson are considered to be proportional to execution time. The server remains on RTX A5000; only client times are down-scaled to Jetson. We reported the average time, energy consumption, and CO$_2$ emissions for client ($C$) and server ($S$) in FL and four SL splits (please ref. Section~\ref{perf_ana}) in Table~\ref{tab:singlecol_model_comparison}. 

For \textbf{ResNet-18}, FL (Accuracy 72.34\%) places almost all cost on the client, namely computation time of 133.7 seconds, energy consumption of $\approx 20$kJ, and 2.6g of CO$_2$ emission, while server cost is negligible (Table~\ref{tab:singlecol_model_comparison}(a)). However, SL$_{15,85}$ that computes only 15\% of layers on farm devices and slashes client time to 14 seconds (89.5\% reduction from FL) at the expense of raising client energy in the range $\left[31.23, 42.79\right]$kJ and CO$_2$ to $\approx 5$g and increasing server use to $0.393$ seconds, energy to $0.086$kJ, and CO$_2$ emission to $0.01135$g. Intermediate splits such as SL$_{40,60}$ achieved 75.98\% accuracy and report client time as 35 seconds and server time as 0.241 seconds, with moderate energy consumption and CO$_2$ emission trade-offs.

\begin{table}[h]
\footnotesize
\setlength{\tabcolsep}{1pt}
\renewcommand{\arraystretch}{1.25}
\centering
\caption{Average Time, Energy, and CO$_2$ Emission by Client ($C$) and Server ($S$).}
\vspace{-0.1in}
\begin{subtable}[t]{\columnwidth}
\centering
\caption{ResNet18}
\resizebox{.99\textwidth}{!}{
\begin{tabular}{|l|c|c|c|c|c|c|}
\hline
\textbf{} & \multicolumn{2}{c|}{Time (s)} & \multicolumn{2}{c|}{Energy (kJ)} & \multicolumn{2}{c|}{CO$_2$ (g)} \\
\cline{2-7}
\textbf{Method} & $C$ & $S$ & $C$ & $S$ & $C$ & $S$ \\
\hline
FL & 133.70 $\pm$ 0.04 & 0.137 & 19.65 $\pm$ 1.55 & 0.028 & 2.59 $\pm$ 0.20 & 0.0037 \\
SL$_{75,25}$ & 41.12 $\pm$ 0.53 & 0.204 & 43.28 $\pm$ 11.01 & 0.0649 & 5.71 $\pm$ 1.45 & 0.0086 \\
SL$_{40,60}$ & 34.99 $\pm$ 0.15 & 0.241 & 42.09 $\pm$ 11.33 & 0.0653 & 5.55 $\pm$ 1.49 & 0.0086 \\
SL$_{25,75}$ & 27.91 $\pm$ 0.22 & 0.297 & 47.47 $\pm$ 14.32 & 0.0374 & 6.26 $\pm$ 1.89 & 0.0049 \\
SL$_{15,85}$ & 13.58 $\pm$ 0.43 & 0.393 & 37.01 $\pm$ 5.78 & 0.086 & 4.88 $\pm$ 0.77 & 0.0114 \\
\hline
\end{tabular}
}
\end{subtable}

\vspace{0.2cm}

\begin{subtable}[t]{\columnwidth}
\centering
\caption{GoogleNet}
\resizebox{.99\textwidth}{!}{
\begin{tabular}{|l|c|c|c|c|c|c|}
\hline
\textbf{} & \multicolumn{2}{c|}{Time (s)} & \multicolumn{2}{c|}{Energy (kJ)} & \multicolumn{2}{c|}{CO$_2$ (g)} \\
\cline{2-7}
\textbf{Method} & $C$ & $S$ & $C$ & $S$ & $C$ & $S$ \\
\hline
FL & 194.76 $\pm$ 20.45 & 0.210 & 27.62 $\pm$ 2.74 & 0.038 & 3.65 $\pm$ 0.36 & 0.0050 \\
SL$_{75,25}$ & 69.55 $\pm$ 1.09 & 0.360 & 60.34 $\pm$ 14.49 & 0.217 & 7.96 $\pm$ 1.91 & 0.0286 \\
SL$_{40,60}$ & 56.73 $\pm$ 1.86 & 0.450 & 61.34 $\pm$ 17.30 & 0.214 & 8.09 $\pm$ 2.28 & 0.0283 \\
SL$_{25,75}$ & 52.19 $\pm$ 2.14 & 0.480 & 60.97 $\pm$ 14.81 & 0.227 & 8.04 $\pm$ 1.95 & 0.0300 \\
SL$_{15,85}$ & 39.04 $\pm$ 1.73 & 0.560 & 63.08 $\pm$ 21.72 & 0.235 & 8.32 $\pm$ 2.87 & 0.0311 \\
\hline
\end{tabular}
}
\end{subtable}

\vspace{0.2cm}

\begin{subtable}[t]{\columnwidth}
\centering
\caption{MobileNet}
\resizebox{.99\textwidth}{!}{
\begin{tabular}{|l|c|c|c|c|c|c|}
\hline
\textbf{} & \multicolumn{2}{c|}{Time (s)} & \multicolumn{2}{c|}{Energy (kJ)} & \multicolumn{2}{c|}{CO$_2$ (g)} \\
\cline{2-7}
\textbf{Method} & $C$ & $S$ & $C$ & $S$ & $C$ & $S$ \\
\hline
FL & 196.01 $\pm$ 0.32 & 0.167 & 32.96 $\pm$ 1.68 & 0.033 & 0.43 $\pm$ 0.02 & 0.0043 \\
SL$_{75,25}$ & 65.10 $\pm$ 0.06 & 0.260 & 7.93 $\pm$ 1.84 & 0.039 & 1.05 $\pm$ 0.24 & 0.0051 \\
SL$_{40,60}$ & 51.95 $\pm$ 0.06 & 0.333 & 6.80 $\pm$ 1.93 & 0.048 & 0.90 $\pm$ 0.25 & 0.0065 \\
SL$_{25,75}$ & 42.68 $\pm$ 0.04 & 0.360 & 7.00 $\pm$ 1.78 & 0.035 & 0.92 $\pm$ 0.23 & 0.0046 \\
SL$_{15,85}$ & 26.50 $\pm$ 0.03 & 0.493 & 4.50 $\pm$ 1.02 & 0.060 & 0.59 $\pm$ 0.13 & 0.0079 \\
\hline
\end{tabular}
}
\end{subtable}
\label{tab:singlecol_model_comparison}
\vspace{-0.1in}
\end{table}

For \textbf{GoogleNet}, FL achieved 63.15\% accuracy and again burdens the farm devices with computation time $(195\pm21)$ seconds, energy consumption in the range of $\left[24.9, 30.4\right]$, CO$_2$ emission of $\approx 4$ g, and barely taxes the server (Table~\ref{tab:singlecol_model_comparison}(b)). The SL$_{25,75}$ split achieved an accuracy of 80.35\%; however, the farm device/client latency falls to $52.19\pm2.14$ seconds, client energy to $60.97\pm14.81$kJ, and CO$_2$ to $10$g (approx), while server costs double relative to the FL approach. Even SL$_{40,60}$ (accuracy = 78.16\%) took computational time of around 58 seconds and consumes the median energy of $61.34$kJ. 

For \textbf{MobileNetV2} architecture, the SL$_{15,85}$ split maintains 80.98\% accuracy while reducing client time to $26.50$ seconds and energy to $4.50\pm1.02$kJ (CO$_2$ = 0.59 g), with server time and energy consumption rising to $0.493$ seconds and $0.0596$kJ respectively, as shown in Table~\ref{tab:singlecol_model_comparison} (c). Another split, SL$_{25,75}$, achieved an accuracy of 82.35\% and further cuts on-device computational time and CO$_2$, demonstrating that modest offloading over the cloud can even improve accuracy while drastically lowering edge resource demands.

We observed that SL consistently reduces execution time on the client side across all models; however, it is not directly proportional to reductions in energy consumption and CO$_2$ emissions. In ResNet18 and GoogleNet, for example, client time drops significantly with SL splits while the energy/emissions increase in some cases than FL. This counterintuitive behavior arises because SL introduces multiple forward and backward passes at shallower layers that can be computationally inefficient. These layers often involve high-resolution feature maps and large memory footprints, which incur greater energy costs even for short execution times. Moreover, frequent data exchange with UAV also increases the energy overhead. In contrast, MobileNetV2, with its lightweight inverted residual structure, handles early-layer computations more efficiently, allowing SL to reduce both time and energy simultaneously. This highlights that model architecture is crucial in determining whether SL yields energy savings or trade-offs.

\section{Conclusion}~\label{conclusion}
\vspace{-0.1in}

In this paper, we introduced \textit{eEnergy-Split}, an energy-aware framework that combines split learning (SL) with an optimal sensor deployment strategy and a UAV path planning method based on an exact TSP solver. The proposed framework enables efficient and privacy-preserving collaborative learning across edge devices in smart farming environments. By partial offloading to the server, \textit{eEnergy-Split} significantly reduces on-device energy consumption and improves model accuracy compared to FL. Extensive experiments on the pest dataset demonstrate that SL improves classification performance while lowering UAV energy consumption, supporting longer mission durations. Notably, our findings reveal that the energy efficiency of SL is model-dependent. MobileNetV2 achieves significant reductions in energy use and $CO_2$ emissions compared to ResNet18 and GoogleNet. These results emphasize the importance of aligning model architecture with system-level design choices to maximize the benefits of SL in resource-constrained environments. Future research will focus on reducing communication overhead in SL through activation compression and sparsification techniques. We also plan to explore adaptive split point selection based on real-time energy profiling and network conditions. Furthermore, we intend to develop a full hardware prototype and deploy it in an actual farm environment, creating a real-world testbed to conduct live experiments and validate the framework’s performance through real-time analysis under practical conditions.


\vspace{2pt}
\noindent \textbf{Acknowledgments:} This work is supported by the NSF award \#s 2331554 (I-Corps), SCC-1952045 (SIRAC), PFI-RP-2431990, AI-ENGAGE-2520346 (HARVEST).

\bibliographystyle{IEEEtran}
\bibliography{MASS}


\end{document}